\documentclass{article}
\usepackage{spconf,amsmath,graphicx}
\usepackage{hyperref}
\usepackage{amssymb}

\title{Complex Ratio Masking for Singing Voice Separation}
\name{Yixuan~Zhang\textsuperscript{1},
      Yuzhou~Liu\textsuperscript{1} and
      DeLiang~Wang\textsuperscript{1,2}
    \thanks{This work was supported in part by an NIDCD grant (R01DC012048) and by the Ohio Supercomputer Center.}}
\address{\textsuperscript{1}Department of Computer Science and Engineering, The Ohio State University, USA\\
\textsuperscript{2}Center for Cognitive and Brain Sciences, The Ohio State University, USA\\
{\normalsize{\href{mailto:zhang.7388@osu.edu, liu.2376@osu.edu,wang.77@osu.edu }{\{zhang.7388, liu.2376, wang.77\}@osu.edu}}}}

\begin{document}
\ninept
\maketitle
\begin{abstract}
Music source separation is important for applications such as karaoke and remixing. Much of previous research focuses on estimating short-time Fourier transform (STFT) magnitude and discarding phase information. We observe that, for singing voice separation, phase can make considerable improvement in separation quality. This paper proposes a complex ratio masking method for voice and accompaniment separation. The proposed method employs DenseUNet with self attention to estimate the real and imaginary components of STFT for each sound source. A simple ensemble technique is introduced to further improve separation performance. Evaluation results demonstrate that the proposed method outperforms recent state-of-the-art models for both separated voice and accompaniment.
\end{abstract}
\section{Introduction}
\label{sec:intro}

Music source separation is the task of separating or isolating different sound sources (or components) from a music recording. Music components can be voice, piano, violin and other accompaniments. As an important task of music source separation, singing voice separation has received a lot of attention due to commercial applications such as automatic karaoke creation and lyric generation, and its role in facilitating related tasks such as singing melody extraction and singer identification \cite{li2007separation}, \cite{rafii2018overview}. Traditionally, matrix decomposition methods such as non-negative matrix factorization (NMF) \cite{ozerov2009multichannel}, sparse coding \cite{blumensath2004unsupervised} and independent component analysis \cite{jang2003maximum}, are used to address music source separation. These methods model a mixture signal as a weighted sum of bases under certain assumptions. However, the diversity of music signals makes such assumptions (e.g. statistical independence) difficult to hold \cite{rafii2018overview}. 

With the rapid development of deep learning and the availability of large databases, deep neural network (DNN) based methods \cite{nugraha2016multichannel}, \cite{uhlich2017improving}, \cite{stoller2018adversarial} have achieved substantial improvements over conventional methods for singing voice separation. The general approach is to perform source separation in the frequency domain by feeding spectral features calculated from an audio mixture to DNN. It is commonly assumed that the magnitude spectrogram carries sufficient information for source separation, and the phase spectrogram is not considered \cite{takahashi2017multi}, \cite{takahashi2018mmdenselstm}, \cite{yuzhou2019attention}. In other words, only magnitude spectrogram is estimated for each source and the separated audio is re-synthesized with mixture phase. Although this assumption is long held in speech enhancement \cite{wang1982unimportance}, we observe that phase is important for singing voice separation. 

After Williamson et al. \cite{williamson2015complex} first studied complex-domain speech separation by introducing the real-imaginary representation instead of the standard magnitude-phase representation,  several studies attempt to estimate the phase of music sources. In \cite{magron2018reducing}, methods such as Wiener filter and iterative procedure that incorporate phase constraints are discussed in singing voice separation systems. Lee et al. \cite{lee2017fully} estimate the complex-valued STFT of music sources by a complex-valued deep neural network. PhaseNet \cite{takahashi2018phasenet}  handles phase estimation as a classification problem. Moreover, there are recent studies \cite{stoller2018wave}, \cite{lluis2018end}, \cite{defossez2019music} that address the music source separation problem in the waveform domain. Wave-U-Net \cite{stoller2018wave} adapts the U-Net structure to the time domain, although its performance is worse than the best spectrogram-based published in SiSEC 2018 \cite{stoter20182018} (\cite{lluis2018end}). Very recently, a time-domain model \cite{defossez2019music} achieves comparable performance with state-of-the-art spectrogram-based models for extracting instrument sources. But their performance for singing voice separation is still lower than that in \cite{takahashi2018mmdenselstm}. 

In contrast to the above studies, we estimate the magnitude and phase spectrograms simultaneously in the complex domain. Our work is inspired by complex-domain speech separation \cite{williamson2015complex}. After comparing several training targets in the complex domain, such as the complex ideal ratio mask (cIRM) \cite{williamson2015complex} and the target complex spectrum \cite{lee2017fully}, \cite{tan2019complex}, we estimate the cIRM but define the loss function in terms of the complex spectrogram. Our DNN structure is based on the self-attention Dense-UNet (SA-DenseUNet) in \cite{yuzhou2019attention}, which reports the state-of-the-art results on the basis of magnitude spectrogram. In addition, we adopt an ensemble learning strategy \cite{zhang2016deep} to boost the performance. Our method substantially outperforms current state-of-the-art approaches.

The rest of this paper is organized as follows. In the next section, we describe the details of our proposed method. The experiment setup and evaluation results are described in Section 3. Concluding remarks are given in Section 4.

\section{Proposed Method}
\label{sec:method}

\subsection{Importance of phase in singing voice separation}
\begin{figure}
  \begin{center}
  \includegraphics[width=3.39in]{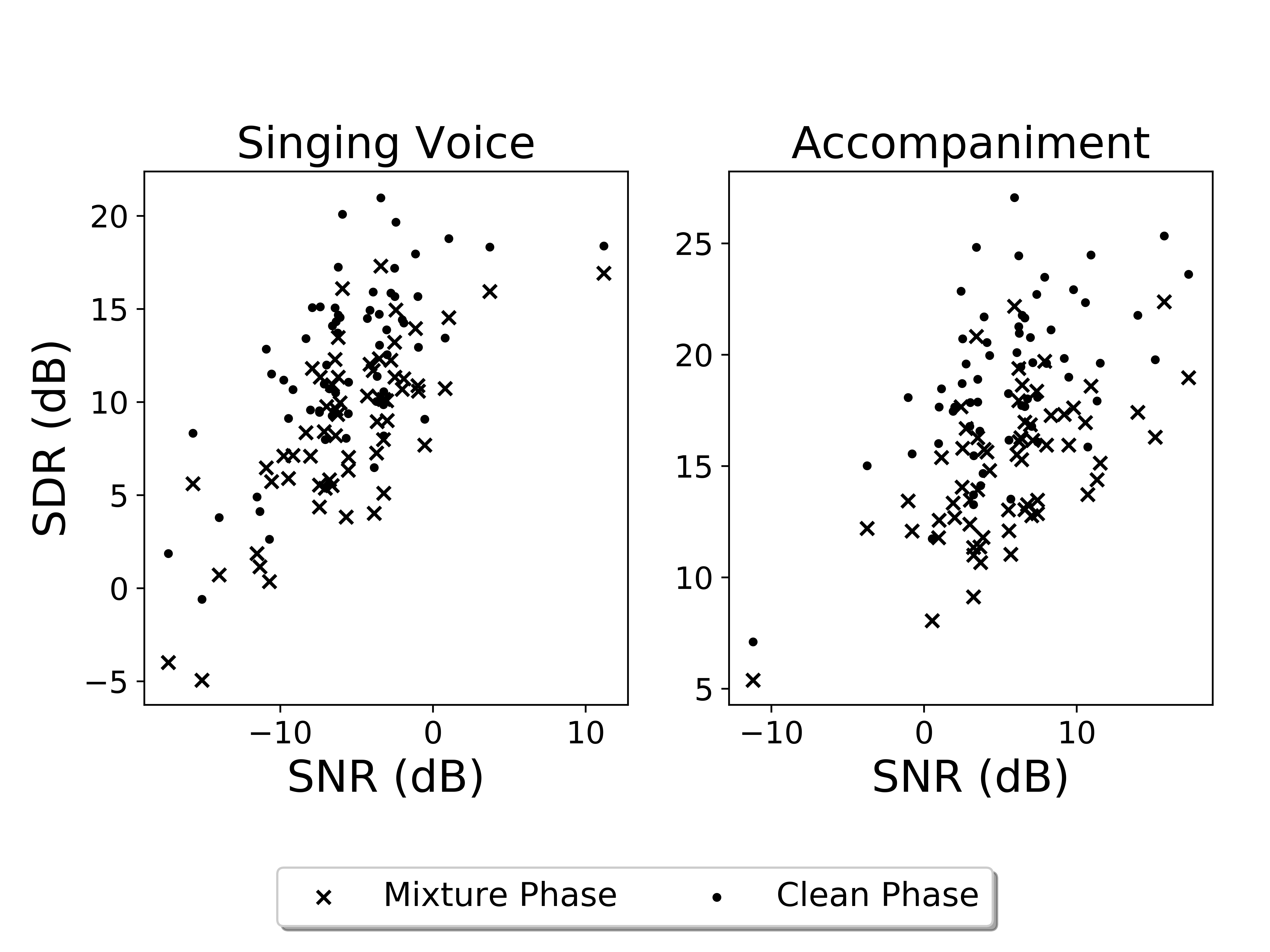}\\
  \caption{SDR of separated singing voice (left) and accompaniment (right) with mixture phase versus clean phase.}\label{comparison_clean_phase}
  \end{center}
\end{figure}

To examine the the importance of phase, we use SA-DenseUNet \cite{yuzhou2019attention} to estimate magnitude spectrograms of singing voice and accompaniment for 63 songs with different signal-to-noise ratios (SNRs) from the test set described in Section 3.1, and compare the signal-to-distortion ratio (SDR) of output audios re-synthesized with clean phase versus mixture phase. Fig. \ref{comparison_clean_phase} shows the experimental results for singing voice and accompaniment respectively. Each song at a certain SNR has two scores on a vertical line in each plot: cross `x' and dot `.', which represent output SDR with mixture phase and clean phase respectively. For both singing voice and accompaniment, the use of clean phase leads to considerable improvement, about 4 to 5 dB on average. Note that we calculate the SNR of singing voice/accompaniment of each song without removing non-vocal portions of a mixture, which do not constitute a large proportion. These results demonstrate the importance of phase in singing voice separation, and motivate us to investigate complex-domain separation.

\subsection{Training targets}

Proper training targets play a significant role in supervised source separation \cite{wang2018supervised}. Mapping-based methods directly estimate the spectra of singing voice and accompaniment from the spectra of the mixture. For complex spectral mapping, the training target is the real and imaginary components of the STFT of a target source \cite{williamson2015complex} \cite{lee2017fully} \cite{tan2019complex}. 

On the other hand, masking-based methods estimate the time-frequency mask of a target source. We propose to employ the cIRM of a target source as the training target. With this target one can define the loss function as the difference between the cIRM and its estimate (cRM). Another way is to define the loss as the difference between clean spectrum and estimated spectrum \cite{liu2019divide}:

{
\footnotesize
\begin{align}
    L =  \sum_{j = 1,2} \bigg[|Re(S_j - cRM_j \odot Y))| 
    + |Im(S_j - cRM_j \odot  Y))| \bigg],
\end{align}
}%
where $cRM_j$ is an estimate of the cIRM for source $j$, and $\odot$ denotes element-wise complex multiplication. $Y$ denotes the complex STFT of the input mixture, and $S_1$ and $S_2$ represent the complex STFT of singing voice and accompaniment respectively. We make use of the loss in (1) along with complex ratio masking.

\subsection{Complex SA-DenseUNet}
\begin{figure*}[!t]
\centering
  \includegraphics[width=6.8in,,height=1.3in]{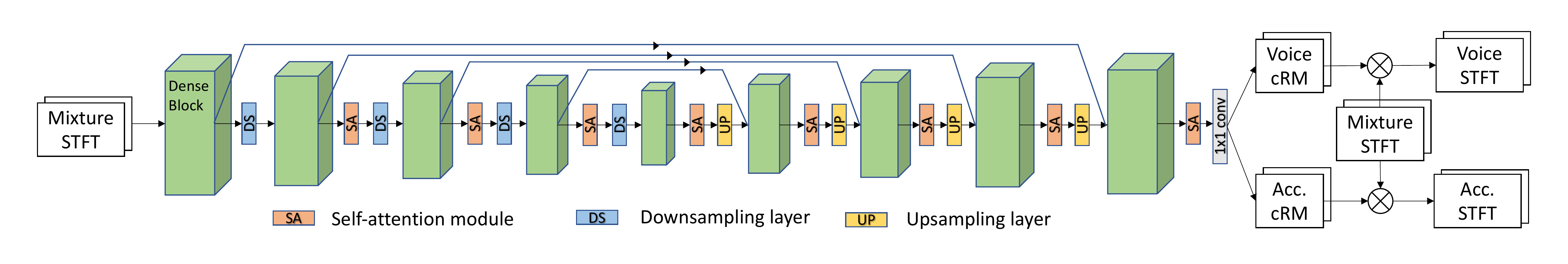}\\
  \caption{Diagram of network structure. 
  The input, masks, outputs are all defined in the complex domain. Acc. refers to accompaniment.}\label{network_structure}
\end{figure*}

This study extends SA-DenseUNet \cite{yuzhou2019attention} to estimate the real and imaginary components of the STFT of target sources. The network diagram is shown in Fig. \ref{network_structure}. The input has three dimensions: frequency, time and channel, with the real and imaginary components treated as two separate channels. The network adopts a DenseUNet structure which consists of an encoder and a corresponding decoder. The encoder comprises a series of densely-connected convolutional layers (referred to as dense blocks), self-attention modules and downsampling layers. The encoder enables the network to generate higher level features, making it possible for the network to process longer temporal contexts efficiently. The decoder consists of dense blocks, self-attention modules and upsampling layers. Through the decoder, the resolution of encoded features is increased to their original levels. Skip connections are introduced between the encoder and the decoder to connect two dense blocks with the same scale, and they transmit relatively raw features from earlier layers to later ones. 
A dense block contains several convolutional layers, and the input of each layer is a concatenation of the outputs from all preceding layers. The output of the $i$th layer $x_i$ can be formulated as
\begin{align}
    x_i = H_i([x_{i-1}, x_{i-2}, ..., x_0]),
\end{align}
where the dense block input is denoted as $x_0$ and $H_i(\cdot)$ denotes nonlinear transformation in the $i$th layer which, in our case, is a convolutional layer followed by ELU (exponential linear unit) activation function. Symbol $[\cdot]$ denotes the concatenation operation. A dense block allows each layer to receive and reuse the output features from preceding layers, which improves parameter efficiency  and reduces redundant calculations.

In addition, several self-attention modules are placed after dense blocks at different levels. Self-attention is designed to capture the repetitive patterns in the accompaniment. To compute Key $\mathbf{K}$, Query $\mathbf{Q}$ and Value $\mathbf{V}$ matrices, the input of a self-attention module, with dimensions $F \times T \times C$, is first fed into three $1 \times 1$ convolutional layers. $F$, $T$, $C$ denote frequency, time and channel dimensionality, respectively. The first two layers reduce $C$ to $C'$ and the last one has the original number of channels $C$. The outputs of the first two convolutional layers are reshaped to 2-D representations with dimensions $(C^\prime \cdot F) \times T$, and that of the last one with dimensions $(C \cdot F) \times T$. The first two feature maps are further processed to $\mathbf{K}$, $\mathbf{Q}$ by reducing the size of the first dimension from $(C' \cdot F)$ to $E$. The last feature map is denoted as $\mathbf{V}$. With $\mathbf{Q} \in \mathbb{R}^{E \times T}$ and $\mathbf{K} \in \mathbb{R}^{E \times T}$ matrices, the attention map $\boldsymbol{\beta} \in \mathbb{R}^{T \times T}$ is calculated by:
\begin{align}
    \beta_{i,j} = \frac{\exp{a_{ij}}}{\sum^{T}_{j=1} exp(a_{ij})},
\end{align}

where $a_{ij}=\mathbf{Q}(i)^T\mathbf{K}(j)$ and $\beta_{i,j}$ denotes the value of the $i$th row and the $j$th column of attention map $\boldsymbol{\beta}$. Note that $\mathbf{Q}(t) \in \mathbb{R}^{E \times 1}$ and $\mathbf{K}(t) \in \mathbb{R}^{E \times 1}$ represent the time segment of Query and Key. The attention map matrix is then multiplied with the Value matrix $\mathbf{V} \in \mathbb{R}^{(C \cdot F) \times T}$ to get the output, which is further reshaped to the original size of the input and concatenated with the input feature to feed the next level. In this process, voice and accompaniment segments attract attention separately at different levels of SA-DenseUNet. 

Finally, the neural network estimates the real and imaginary components of the cIRM of each source, which are then multiplied with the complex STFT of the mixture audio to get the estimated complex STFT of each source
 \cite{williamson2015complex}.

\subsection{Multi-context averaging}

Multi-context averaging is an ensemble learning technique that averages the outputs of DNNs whose inputs have different contexts \cite{zhang2016deep}. Here we exploit different contexts of the input by using different window lengths. Suppose that $P$ $(P >1)$ DNNs form a DNN ensemble and every DNN has a different window length. Given the mixture $Y$ and the corresponding clean singing voice $S_1$ and accompaniment $S_2$, the $p$th DNN is trained to estimate the real and imaginary components of the STFT of $S_1$ and $S_2$ with window length set to $w_p$.
In the test stage, for the $p$th DNN, the corresponding complex STFT is also computed with window length $w_p$. The outputs of all $P$ DNNs are transformed to the time domain by inverse STFT. After that, we calculate the average of waveform outputs by
\begin{align}
\hat{s}_j = \frac{1}{P} \cdot \sum_{p=1}^P \hat{s}_{j,p},
\end{align}
where $\hat{s}_j$ corresponds to the final estimated singing voice for $j = 1$ and accompaniment for $j = 2$. $\hat{s}_{j,p}$ corresponds to the estimated sound waveform from the $p$th DNN. 

\subsection{Network details}

In complex SA-DenseUNet, for each source we use two output layers with linear activation to estimate the real and imaginary components of the cIRM with the value range of $(-\infty, +\infty)$ as illustrated in Fig. 2. A $1\times1$ convolutional layer is applied before output layers for feature reorganization. As portrayed in Fig. 2, our model has in total 9 dense blocks, 4 downsampling layers and 4 upsampling layers. Each dense block has 4 convolutional layers with kernel size of 3, stride of 1, and use ELU as the activation function. In a dense block, the first three layers use SAME padding and the last layer uses VALID padding. All input audios are downsampled to 16 kHz. STFT is computed with 64 ms frame length and 25\% frame shift for experiments in Section 3.2. The network input corresponds to an excerpt of a song which contains 1250 time frames. All networks are trained using the ADAM \cite{kingma2014adam} optimizer with a learning rate of 0.00005 for 35 epochs. The model with the best loss on the validation set is used for testing. Test songs are divided into 20s excerpts with $3/4$ overlap between consecutive excerpts.

\section{Evaluation and Comparison}
\label{sec:pagestyle}

\subsection{Experimental setup}
To facilitate comparisons, our evaluations are set up in a similar way to \cite{yuzhou2019attention}. The dataset is constructed from DSD100 \cite{liutkus20172016}, MedleyDB \cite{bittner2014medleydb} and CCMixter\footnote{www.ccmixter.org}. DSD100 consists of Dev and Test sets, and each contains 50 songs. MedleyDB includes 122 songs in which 70 tracks contain vocals. CCMixter has 50 vocal songs. The training set contains 450 songs, in which 50 are from DSD100's Dev Set and 400 songs are generated by randomly scaling, shifting, remixing different music sources from these 50 songs for data augmentation. The validation set contains one third of tracks from MedleyDB and CCMixter and half of tracks from DSD100's Test Set. The test set is constructed with another third of tracks from MedleyDB and CCMixter, and the remaining half of DSD100's Test Set. Since not all tracks in MedleyDB contain vocals, we remove the non-vocal tracks from the test set, which results in 20 vocal tracks from MedleyDB. The singing voice SNR of the test set is -5.6 dB on average. All songs are downsampled from 44.1 kHz to 16 kHz for training and testing, to avoid high computational costs. For evaluation metric, we use \textit{mir\_eval} \cite{raffel2014mir_eval} to calculate the average SDR, SIR, SAR of each song. 

\subsection{Comparisons of different training targets}

Table \ref{targets} presents the evaluation results for different training targets described in Section 2.2. 
We compare a magnitude-domain and two complex-domain training objectives. One complex-domain training target is the target complex spectrum, referred to as the TCS \cite{tan2019complex}. The other is the cIRM with the loss function defined in complex spectrum, referred to as the cIRM-CS. It can be observed that complex-domain training targets achieve better SDR and SIR performances compared with the target magnitude spectrum (TMS) \cite{wang2018supervised}. The proposed cIRM-CS performs uniformly better than the TCS, and achieves the best SDR score. 

With the estimates of real and imaginary components, phase can be easily estimated. To evaluate the effectiveness of the estimated phase, we take the magnitude STFT estimated by SA-DenseUNet and re-synthesize the source waveform with either mixture phase or the phase estimated by cRM. As shown in Table \ref{magnitude-phase}, with the estimated phase, the SDRs of singing voice and accompaniment are increased by 0.69 dB and 0.89 dB respectively, demonstrating the effectiveness of phase estimation. 

\begin{table}[!t]
\renewcommand{\arraystretch}{1.4}
\caption{Comparison of different training targets}
\label{targets}
\centering
\small
\begin{tabular}{l|lll|lll}
\hline
 & \multicolumn{3}{c|}{Singing Voice}                                                 &  \multicolumn{3}{c}{Accompaniment}                            \\ \cline{2-7} 
Metric (dB) & \multicolumn{1}{c}{SDR} & \multicolumn{1}{c}{SIR} & \multicolumn{1}{c|}{SAR} & \multicolumn{1}{c}{SDR} & \multicolumn{1}{c}{SIR} & \multicolumn{1}{c}{SAR} \\ \hline
 TMS &         8.08              &           15.44            &  9.34            &         14.10       &           18.42           &           16.50      \\ 
TCS &   8.40                    &     18.74                  &      9.07                 &          14.50             &       22.29                & 15.70 \\
 cIRM-CS &        8.92               &           20.04            &       9.54                &       14.99                &              23.06         & 16.12 \\ \hline
\end{tabular}
\end{table}

\begin{table}[!t]
\renewcommand{\arraystretch}{1.4}
\caption{Evaluation of phase estimation}
\label{magnitude-phase}
\centering
\footnotesize
\begin{tabular}{l|lll|lll}
\hline
  & \multicolumn{3}{c|}{Singing Voice}                                                 &  \multicolumn{3}{c}{Accompaniment}                            \\ \cline{2-7} 
 Metric (dB) & \multicolumn{1}{c}{SDR} & \multicolumn{1}{c}{SIR} & \multicolumn{1}{c|}{SAR} & \multicolumn{1}{c}{SDR} & \multicolumn{1}{c}{SIR} & \multicolumn{1}{c}{SAR} \\ \hline
 Mixture phase  &         8.08              &           15.44            &  9.34            &         14.10       &           18.42           &           16.50     \\
 Estimated phase &   8.77                    &     19.52                  &      9.42                 &          14.99             &            22.67           & 16.23 \\ \hline
\end{tabular}
\end{table}

\subsection{Multi-context averaging}

\begin{figure}
  \begin{center}
    \includegraphics[width=3.35in,height=1.5in]{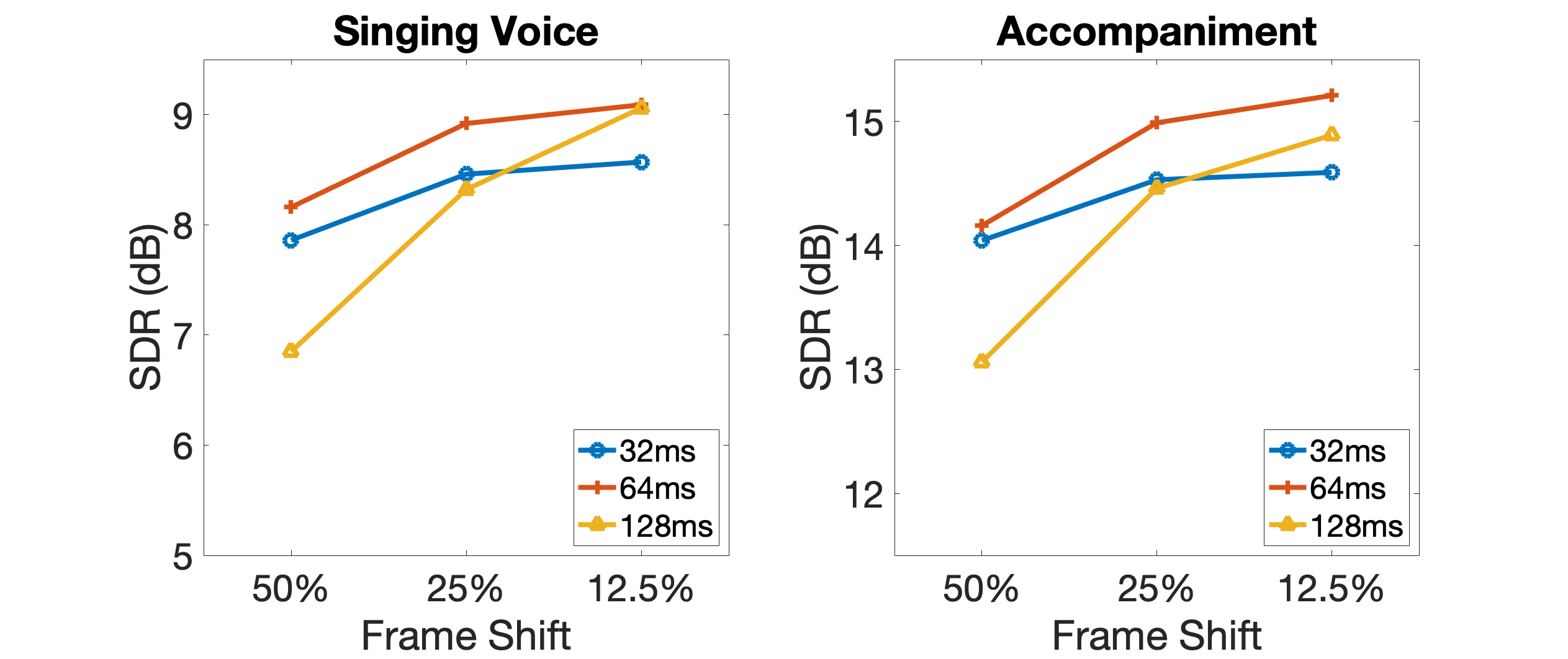}
  \caption{Average SDR (dB) results for different frame lengths and frame shifts.}
  \label{diff_input_context}
  \end{center}
\end{figure}

Since the cIRM-CS outperforms other training targets, we use it in the following experiments. To create inputs with different contexts, we first train our complex SA-DenseUNet with different window lengths and shifts. We test three window lengths, i.e. 32 ms, 64 ms, 128 ms, and for each window length, three frame shifts: 12.5\%, 25\%, 50\%. In total, 9 experimental results are obtained 
and plotted in Figure \ref{diff_input_context}. We can observe that with the same window length, the performance improves with reduced frame shift for both singing voice and accompaniment. With a smaller frame shift, each time sample is estimated more times, leading to a larger ensemble and better performance. As observed in Figure \ref{diff_input_context}, the model achieves the best result with the 64 ms window length and 12.5\% frame shift. 

We choose 3 complex SA-DenseUNets whose window lengths of 32 ms, 64 ms, 128 ms to create a multi-context averaging (MCA) ensemble, all with 12.5\% frame shift. Table \ref{multicontext} compares the results of the MCA ensemble and the single best window length of 64 ms with 12.5\% frame shift. Compared to the single best window length, the MCA network improves the mean SDRs of singing voice and accompaniment by 0.64 dB and 0.36 dB respectively.

\begin{table}[!t]
\scriptsize
\renewcommand{\arraystretch}{1.4}
\caption{Average test results on the test set}
\label{multicontext}
\centering
\begin{tabular}{l|lll|lll}
\hline
 & \multicolumn{3}{c|}{Singing Voice}                                                 &  \multicolumn{3}{c}{Accompaniment}                            \\ \cline{2-7} 
 Metric (dB) & \multicolumn{1}{c}{SDR} & \multicolumn{1}{c}{SIR} & \multicolumn{1}{c|}{SAR} & \multicolumn{1}{c}{SDR} & \multicolumn{1}{c}{SIR} & \multicolumn{1}{c}{SAR} \\ \hline
Complex SA-DenseUNet &         9.09              &           20.15            &  9.76            &         15.21       &           \textbf{22.77}           &           16.45 \\ 
 Multi-context averaging &        \textbf{9.73}               &       \textbf{20.76}            &       \textbf{10.36}                &       \textbf{15.57}                &              21.78         & \textbf{17.28} \\ \hline
\end{tabular}
\end{table}

\subsection{Overall evaluation and comparisons with other methods}

\begin{table}[!t]
\renewcommand{\arraystretch}{1.4}
\caption{Comparison of median SDR values on DSD100 dataset}
\label{state-of-the-art}
\small
\centering
\begin{tabular}{l|l|l}
\hline
Metric (dB)                & \multicolumn{1}{c|}{Singing Voice} & \multicolumn{1}{c}{Accompaniment} \\ \hline \hline
MMDenseNet \cite{takahashi2017multi}                &   6.00                            &  12.10                     \\
MMDenseLSTM \cite{takahashi2018mmdenselstm}            & 6.31                               & 12.73                     \\
SA-SHN-4 \cite{yuan2019skip}                  & 6.44                                    &    12.60              \\
SA-DenseUNet \cite{yuzhou2019attention} &    7.72                          &  13.90                      \\
Proposed                      & \textbf{9.78}                     & \textbf{15.20}             \\ \hline
\end{tabular}
\end{table}

Finally, we evaluate the proposed method and compare with four state-of-the-art methods, which are MMDenseNet \cite{takahashi2017multi}, MMDenseLSTM \cite{takahashi2018mmdenselstm}, and SA-SHN-4 \cite{yuan2019skip}, SA-DenseUNet \cite{yuzhou2019attention}. MMDenseNet has a similar encoder-decoder structure to SA-DenseUNet. The network is applied to multiple frequency bands to learn local patterns. MMDenseNetLSTM further improves MMDenseNet by incorporating BLSTM  (bidirectional long short-term memory) layers to learn longer contexts. In the SiSec2018 campaign, both MMDenseNet and MMDenseLSTM achieved very good results in singing voice separation and MMDenseLSTM held the highest SDR score for separating singing voice and accompaniment. SA-SHN-4, an attention-driven network published very recently also achieves competitive results in singing voice separation. 

The evaluation and comparisons are conducted on the
DSD100 Test Set, which is used in the 2016 signal separation evaluation campaign (SiSEC) \cite{liutkus20172016}. To be consistent with the comparison methods, we evaluate the test set at the original sampling frequency of 44.1 kHz. Different from the construction described in Section 3.1, the validation set in this case comprises MedleyDB and CCMixter, and does not include any song from the test set of DSD100. A median SDR score is the median of SDR scores of all songs, and it is documented separately for singing voice and accompaniment. 

The results of the proposed model and comparison systems are presented in Table \ref{state-of-the-art}. From the table one can observe that, among the comparison methods, SA-DenseUNet has the best scores for both singing voice and accompaniment. Our method further outperforms SA-DenseUNet by 2.06 dB for singing voice and 1.30 dB for accompaniment.

\section{Concluding Remarks}
This study addresses complex-domain deep learning for singing voice separation. We observe that phase is important for singing voice separation. We find the cIRM to be an effective training target when the loss is defined in terms of complex spectrogram. Moreover, we introduce a simple ensemble learning technique. Systematic evaluation results show that the proposed method produces outstanding separation results, outperforming current state-of-the-art methods.

\clearpage

\begingroup
\newif\ifgobblecomma
\gobblecommafalse 
\edef\FZ{?}
\edef\KM{,}
\catcode`?=\active
\catcode`,=\active
\def?{\FZ\gobblecommatrue}
\def,{\ifgobblecomma\gobblecommafalse\else\KM\fi}
\bibliographystyle{IEEEbib}
\bibliography{strings,refs}
\endgroup


\end{document}